\documentclass[aps,prl,twocolumn,groupedaddress]{revtex4}
\usepackage{graphics,epsfig}
\begin{document}

%\draft
\title{Theory of enhanced performance emerging in a \\
sparsely-connected competitive population}
\author{T.S. Lo$^{1}$, K.P Chan$^{1}$, P.M. Hui$^{1}$, and N.F. Johnson$^{2}$}
\address{$^{1}$Department of Physics, The Chinese University of Hong Kong, Shatin,
New Territories, Hong Kong \\
$^{2}$Physics Department, University of Oxford, Clarendon Laboratory,
Oxford OX1 3PU, U.K.}

\date{\today}

\begin{abstract} We provide an analytic theory to explain Anghel
{\em et al.}'s recent numerical finding whereby a maximum in the
global performance emerges for a sparsely-connected competitive
population [Phys. Rev. Lett. {\bf 92}, 058701 (2004)].  We show
that the effect originates in the highly-correlated dynamics of
strategy choice, and can be significantly enhanced using a simple
modification to the model. \noindent{PACS numbers: 02.50.Le,
05.65.+b, 87.23.Ge, 89.65.Gh} \vskip0.2in

\end{abstract}

%\begin{abstract}
%We provide an analytic theory to explain the numerical finding of
%Anghel et al. [Phys. Rev. Lett. {\bf 92}, 058701 (2004)]
%concerning global fluctuations in a multi-agent model with finite
%interconnectivity. In particular, we explain the surprising
%maximum in global performance which emerges at small connectivity
%and with small agent memory. Our theory centers around the highly
%correlated, non-random dynamics which develop within the strategy
%pool. We also show how the global performance can be significantly
%enhanced using a simple modification to Anghel {\em et al.}'s
%model. \vskip0.1in \noindent{PACS numbers: 02.50.Le, 05.65.+b,
%87.23.Ge, 89.65.Gh} \vskip0.2in
%
%\end{abstract}

\maketitle
%\newpage

There are two particularly active areas of research into Complex
Systems among physicists: multi-agent populations
\cite{arthur1,challet1,johnson2} and complex networks \cite{nets}.
Arthur's bar-attendance problem \cite{arthur1} and its binary
Ising-like simplifications (e.g. the Minority Game (MG)
\cite{challet1,johnson2}) constitute everyday examples of
multi-agent competition for limited resources. However researchers
have only just started considering {\em combining} networks with
such multi-agent systems \cite{anghel,netMG2}. Anghel {\em et al.}
\cite{anghel} reported some fascinating numerical results in which
the fluctuation in the number of agents taking a particular action
can exhibit a {\em minimum} at small connectivity (see Fig.~1
inset). It is truly remarkable that there exists an optimal number
of network connections such that the overall system performance is
maximized, and that this optimal connectivity is actually quite
small.

%%%%%%%%%%%%%%%%%%%%%%%  Fig 1 %%%%%%%%%%%%%%%%%%%%%%%%%%%%%
\begin{figure}
\epsfig{figure=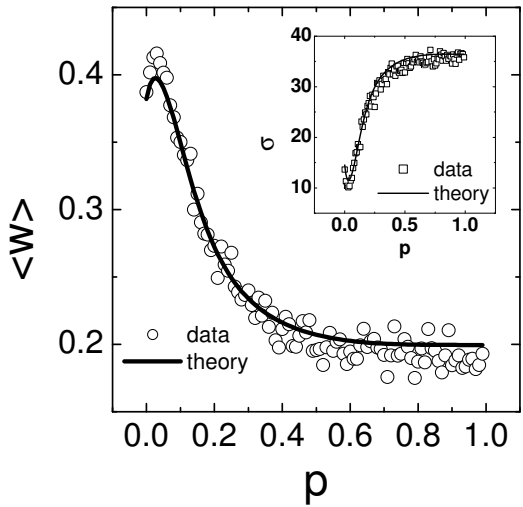,width=\linewidth} \caption{Mean
success rate $\langle w \rangle$ and fluctuation $\sigma$ (inset)
as a function of connectivity $p$ for $m=1$, with $N=101$ and
$s=2$.  Symbols are numerical results. Lines are theoretical
results. Inset shows the minimum in the fluctuation $\sigma$
\cite{anghel}.} \label{figure1}
\end{figure}
%%%%%%%%%%%%%%%%%%%%%%% End fig 1 %%%%%%%%%%%%%%%%%%%%%%%%%%

Anghel {\em et al.}'s results have so far lacked any theoretical
explanation, yet they represent an important challenge for physics
-- not just because of the potential application areas but also
because they expose our limited understanding of complex dynamical
networks. Here we provide the first analytic theory which explains
their remarkable finding. The essential underlying physics
comprises (i) the highly correlated, non-random temporal evolution
of strategy scores, (ii) the tendency to link to future winners
(losers) at low (high) connectivity $p$, and  (iii) the emergence
of different {\em species} of agent characterized by the relative
Hamming distance $D$ of their strategies. The importance of the
underlying dynamics means that approaches based on assumptions of
random histories, e.g. spin-glass theories, are invalid. Our
theory also shows that network connections play a crucial role,
even when only a tiny fraction exist. This enables us to propose a
minor modification to Anghel {\em et al.}'s model which provides
significantly enhanced global performance. Interestingly, there is
recent empirical evidence to suggest that our proposed
`second-best' rule does actually arise in everyday life
\cite{SciAm}. Our theory in the zero-connectivity limit (i.e.
$p=0$) also provides a new microscopic theory for the MG. Note
that the theory we present does not benefit from the
simplifications and hence beauty of conventional many-body theory
in physics. This is because -- in contrast to conventional
physical systems --  the dynamics and configuration space are now
so closely intertwined. However it is {\em precisely} this feature
which makes the problem so interesting for a theoretical physicist
\cite{theory}.

Anghel {\em et al.}'s model \cite{anghel} features $N$ agents who
repeatedly choose between two actions `1' or `0' \cite{challet1}.
The winners are those in the minority group. The global
information is the bit-string containing the $m$ most recent
winning outcomes (i.e. history). Each agent holds $s=2$
strategies.  Each strategy is one of the $2^{2^{m}}$ possible
mappings from the $2^{m}$ histories to action `1' or `0'. All
strategies collect one virtual point (VP) if they predicted the
winning outcome correctly, while each agent collects one (real)
point if he wins. The mean success rate $\langle w \rangle$ is the
average number of real points per agent per turn. The agents are
connected by an undirected random network with $p$ being the
probability that a link between two randomly chosen agents exists.
Each agent compares the cumulated performance of his
best-performing strategy (i.e. his predictor) with that of his
neighbors, and then follows the prediction of whoever holds the
best-performing predictor, including himself. The $p=0$ limit of
the model reduces to the MG. The identity of the best-performing
strategy changes over time, and for $p>0$ the predictor's
performance is generally {\em different} from the agent's
performance. Figure 1 (inset) illustrates the minimum in
fluctuation arising at finite $p$ \cite{anghel}, together with
$\langle w \rangle$ as a function of $p$ for $m=1$. Since these
quantities are simply related, we focus here on $\langle w
\rangle$.

The features of interest occur at small $m$ and small $p$, hence
we focus on the explicit example of $m=1$ (see Fig.~1) and make
the reasonable assumption that the predictors' performance can be
approximated by the $p=0$ results. Generalization to $m=2,3\dots$
and $s>2$ is straightforward but lengthy. For $p=0$ and small $m$,
no single strategy outperforms the others (i.e. no runaway VPs)
and the system restores itself in a finite ($m$-dependent) number
of timesteps. The Eulerian trail acts as a quasi-attractor of the
system's dynamics \cite{eulerian}, yielding anti-persistent
behavior whenever the system revisits a given history node on the
de Bruijn graph of possible history bit-strings.  Let
$\{t_{even}^{\nu}\}$ ($\{t_{odd}^{\nu}\}$) be a set consisting of
the turns in a history series at which a particular history $\nu$
occurred an even (odd) number of times from the beginning of the
run until the moment of the current history $\mu$. For $t \in
\{t^{\mu}_{even}\}$, the agents decide randomly since the strategy
scores are not biased. For $t \in \{t^{\mu}_{odd}\}$, the success
rate is determined by: (i) The number of histories $\kappa$ that
had occurred an odd number of times at the moment of decision.
Since there are $2^{m}$ histories, we have $0 \leq \kappa \leq
2^m$. (ii) The Hamming distance $d$ between an agent's
best-performing strategy and the best performing strategy among
all strategies (BPS) at that particular turn. (iii) The Hamming
distance $D$ between the strategies that an agent holds. For
$s=2$, the probability that the strategies are separated by a
Hamming distance $D$ is given by the binomial coefficient
$C_{D}^{2^{m}}$, where $D = 0,1,\dots,2^{m}$.  For $m=1$, there
are on average $N/4$ agents belonging to the $D=0$ `species' (i.e.
two perfectly correlated strategies), $N/2$ in the $D=1$ `species'
(i.e. two uncorrelated strategies), and $N/4$ in the $D=2$
`species' (i.e. two anticorrelated strategies). For $m=1$, $\kappa
= 0, 1$ or $2$ since there are two possible history bit-strings.
Consider a particular time $t$ corresponding to $\kappa =0$: $t
\in \{t^{\nu}_{even}\}$ for both histories assuming the system
follows the Eulerian trail. Hence the agents become {\em
dynamically segregated} by their performance, according to their
$D$ value. As we now explain, $N/4$ ($D=0$) agents should have a
score of $t/2$, $N/2$ ($D=1$) agents should have a score of
$3t/8$, and $N/4$ ($D=2$) agents should have a score of $5t/16$,
in the long time limit. Prior to a current history of, say, $0$,
each history bit (1 and 0) has occurred an even number of times.
The strategies are all tied. The outcome is thus random (i.e.
coin-toss). Agents with a given $D$ might have won with
probability $w^{(even)}_{\kappa=0} \lesssim 1/2$ or lost with
probability $(1-w^{(even)}_{\kappa=0})$ and hence there are two
subgroups (i.e. won or lost) of agents for a given $D$, with
different sizes. Regardless of the outcome, the system now
corresponds to $\kappa=1$ and the agents' scores can be classified
into six groups. We denote the groups by the label
$\{D,Y\}_{\kappa}$, where $\kappa$ gives the number of history
bit-strings occurring an odd number of times ($0 \leq \kappa \leq
2^{m}$) and $Y$ is the net number of times that the group has won
(i.e., number of winning turns minus number of losing turns)
starting from the most recent occurrence of $\kappa=0$.
%%%%%%%%%%%%%%%%%%%%%%%  Fig 2 %%%%%%%%%%%%%%%%%%%%%%%%%%%%%
\begin{figure}
\epsfig{figure=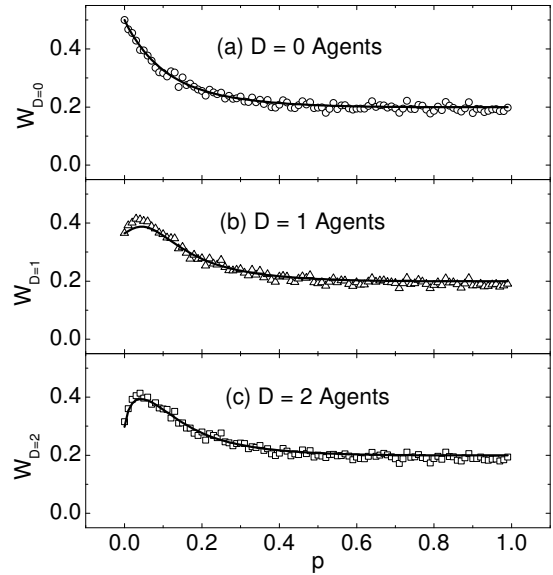,width=\linewidth}
\caption{Numerical (symbols) and theoretical (lines) results for
the average success rate $w_{D}$ of the $D=0,1,2$ agents as a
function of $p$.} \label{figure2}
\end{figure}
%%%%%%%%%%%%%%%%%%%%%%% End fig 2 %%%%%%%%%%%%%%%%%%%%%%%%%%
If the outcome is $0$, then $t \in \{t^{\mu}_{odd}\}$. For the
$D=0$ species, their strategies do not allow them to change their
action and hence the agents who won in the last occurrence
(carrying the label $\{0,1\}_{1}$) will definitely lose and those
who lost (carrying the label $\{0,0\}_{1}$) will definitely win, a
situation denoted by the winning probabilities $S_{\{0,1\}_{1}}=0$
and $S_{\{0,0\}_{1}}=1$. For the $D=1$ species (i.e. two
uncorrelated strategies), those agents who won in the last
occurrence of the history (carrying the label $\{1,1\}_{1}$) must
hold a strategy that points to the most recent winning option, and
hence they will make the same choice -- they will definitely lose
due to the crowd effect. For those who lost, their winning
probability depends on whether their two strategies give the same
or different predictions for the history concerned.  For those
agents with strategies giving the same (different) prediction(s)
(for history $0$ in our example), they will lose (win). Thus the
group of agents labelled by $\{1,0\}_{1}$ will have an average
winning probability of $S_{\{1,0\}_{1}}= 1/2$. For the $D=2$
species, these agents' anti-correlated strategies give different
predictions and hence they will definitely lose, i.e.,
$S_{\{2,1\}_{1}} = S_{\{2,0\}_{1}} = 0$.

If the outcome is $1$ instead of $0$, the situation corresponds to
$\kappa = 1$ since the history 0 has occurred an odd number of
times {\em and} $t \in \{t^{\mu}_{even}\}$ since the current
history $1$ occurred an even number of times.  The strategies' VPs
do not indicate a preference and hence do not lead to a crowd
effect.  In this case, each of the six groups of agents has a
probability of $w^{(even)}_{\kappa=1}$ of winning. As a result,
the population will subsequently be grouped into nine groups
according to the agents' performance in the last two turns, i.e.
win-win, win-lose or lose-win, and lose-lose groups for each value
of $D$. Regardless of the outcome, the system is updated to
$\kappa=2$ and $t \in \{t^{\mu}_{odd}\}$. The instantaneous BPS is
the strategy that predicted correctly the most recent $t \in
\{t^{\mu}_{even}\}$ outcomes for both $\mu=0,1$. The BPS will
predict incorrectly in the following turns, due to the VPs'
anti-persistence. The strategies with the second highest VPs, i.e.
one correct prediction out of two turns, will predict correctly
with probability $1/2$. The momentarily worse-performing strategy
is the one that predicted incorrectly for both histories at $t \in
\{t^{\mu}_{even}\}$. However, it will predict correctly in the
coming $t \in \{t^{\mu}_{odd}\}$ timesteps. Therefore, agents
holding the BPS will use it and are bound to lose. Hence the
$\{D,2\}_{2}$ groups have winning probabilities
$S_{\{D,2\}_{2}}=0$ for $D=0,1,2$, since they hold the BPS. For
the other $D=0$ agents, those who won (lost) in the last
occurrence of the current history will lose (win). Therefore, the
winning probabilities are $S_{\{0,1\}_{2}} = 1/2$ and
$S_{\{0,0\}_{2}}=1$. For the other $D=1$ agents, their winning
probabilities are $S_{\{1,1\}_{2}} = 1/4$ and $S_{\{1,0\}_{2}} =
1/2$. For $D=2$, the $\{2,0\}_{2}$ agents must hold two
anticorrelated strategies of second highest VPs and thus
$S_{\{2,0\}_{2}} = 1/2$.  For the $\{2,1\}_{2}$ group, an agent
may either hold (i) the BPS and the worse-performing strategy, or
(ii) two strategies with the second highest virtual points. For
combination (i), this agent's winning probability is $0$ while for
combination (ii), his winning probability is $1/2$.  Averaging
over these two possibilities gives $S_{\{2,1\}_{2}} = 1/2$. A
common feature of the winning probabilities is that
$\{D,\kappa\}_{\kappa}$ is always zero, i.e. {\em agents with
momentarily high-performance predictors are bound to lose in the
following timesteps.}

This dynamics is valid for  $p\geq 0$. An agent of Hamming
distance $D$ has an average winning probability at $t\in
\{t^{\mu}_{odd}\}$ for a given $\kappa$:
\begin{equation}
w_{D, \kappa}^{(odd)} = \frac{1}{N_{D}}\sum_{y=0}^{\kappa
}N_{\{D,y\}_{\kappa }}S_{\{D,y\}_{\kappa }} , \label{eq1}
\end{equation}
where $N_{D}$ is the number of agents with Hamming distance $D$
and $S_{\{D,y\}_{\kappa }}$ is the winning probability of the
group of agents labelled by $\{D,y\}_{\kappa}$, as discussed
above.  Here $N_{\{D,y\}_{\kappa}}$ is the number of agents in the
group $\{D,y\}_{\kappa}$ which can be found using
$w^{(even)}_{\kappa}$. For small $m$, the probabilities of
occurrence of all histories are equal.  Hence the probability of
having a particular value of $\kappa$ is
$P(\kappa)=C_{\kappa}^{2^{m}}/2^{2^{m}}$. For a given value of
$\kappa$, the probability of having $t \in \{t^{\mu}_{odd}\}$ and
$t \in \{ t^{\mu}_{even}\}$ in a randomly picked turn is $\kappa
/2^{m}$ and $(1 - \kappa/2^{m})$, respectively. Combining with
Eq.(\ref{eq1}), the winning probability of the agents with a given
Hamming distance $D$ is \cite{netMG2}
\begin{equation}
w_{D} = \sum_{\kappa =0}^{2^{m}}P(\kappa ) \left[ \frac{\kappa
}{2^{m}}w_{D, \kappa}^{(odd)} + \left[1 -
\frac{\kappa}{2^{m}}\right] w^{(even)}_{\kappa} \right],
\label{eq2}
\end{equation}
where $w^{(even)}_{\kappa}$ is the winning probability for
timesteps with $t \in \{t_{even}\}$ and can be found by random
walk arguments for $p=0$ and $p\neq 0$ \cite{marketimpact}. For
$p=0$, Eq.~(\ref{eq2}) gives the segregation in success rates
determined by the agents' `species' type $D$. The overall mean
success-rate is hence
\begin{equation}
\left\langle w \right\rangle =\frac{1}{N}\sum_{D=0}^{2^m}w_{D} N_{D}.
\label{eq3}
\end{equation}

%%%%%%%%%%%%%%%%%%%%%%%  Fig 3 %%%%%%%%%%%%%%%%%%%%%%%%%%%%%
\begin{figure}
\epsfig{figure=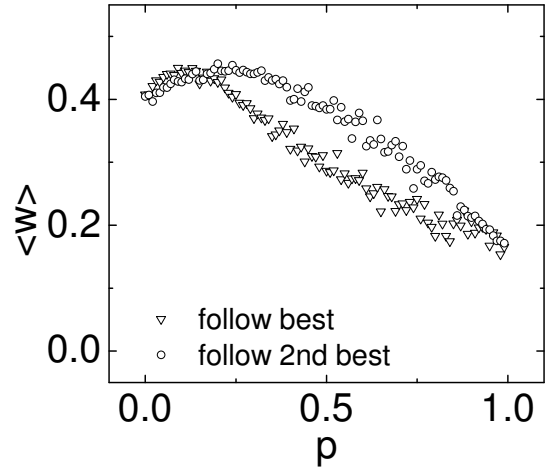,width=\linewidth} \caption{The
mean success rate $\langle w \rangle$ as a function of
connectivity $p$ for our modified model and the model of
Ref.\cite{anghel} for $m=1$.} \label{figure3}
\end{figure}
%%%%%%%%%%%%%%%%%%%%%%% End fig 3 %%%%%%%%%%%%%%%%%%%%%%%%%%

For the range of $p$ where the important features arise, the
agents' predictor performance is identical to the (real) scores or
success rates at $p=0$ discussed above. For $p\neq 0$, the agents
in each group $\{D,y\}_{\kappa}$ can be separated into two
subgroups:
\begin{equation} N_{\{D,y\}_{\kappa }}=\overline{N}_{\{D,y\}_{\kappa
}}+\sum_{D',j}\Delta N_{\{D,y\}_{\kappa },\{D',j\}_{\kappa }}
\label{eq4}
\end{equation} where $\overline{N}_{\{D,y\}_{\kappa }}$ is the number of agents in
the group $\{D,y\}_{\kappa }$ who follow their own predictor, and
$\Delta N_{\{D,y\}_{\kappa },\{D',j\}_{\kappa }}$ is the number of agents who
follow the predictor of a neighbor belonging to group $\{D',j\}_{\kappa }$,
due to the presence of links. Since agents only follow neighbors with better
performing predictors, only links to neighbors in the group with labels
$D'<D$ or $j>y$ (if $D=D'$) are effective. For given $p$, the probability of an
agent in $\{D,y\}_{\kappa}$ having at least one link to agents in
$\{D',j\}_{\kappa }$ is
$1-q^{N_{\{D',j\}_{\kappa }}}$, where $q\equiv 1-p$. The number of agents having
predictor performance better than the group
$\{D,y\}_{\kappa }$ is given by
\begin{equation} A_{\{D,y\}_{\kappa }}=\sum_{D'=0}^{D-1}\sum_{j=0}^{\kappa
}N_{\{i,j\}_{\kappa }}+\sum_{j>y}^{\kappa }N_{\{D,j\}_{\kappa }}.
\label{eq5}
\end{equation} The number of agents $\overline{N}_{\{D,y\}_{\kappa }}$ in
Eq.(\ref{eq4}) is then given by
\begin{equation}
\overline{N}_{\{D,y\}_{\kappa }}=N_{\{D,y\}_{\kappa}}q^{A_{\{D,y\}_{\kappa }}},
\label{eq6}
\end{equation}
since $q^{A_{\{D',j\}_{\kappa }}}$ is the probability of the
agents in group $\{D,y\}_{\kappa }$ not having any links to other
groups with better predictor performance, and so still have
winning probability $S_{\{D,y\}_{\kappa}}$ for $t \in
\{t^{\mu}_{odd}\}$. Agents will follow the prediction of agents in
$\{D',j\}_{\kappa }$ only if (i) they have connections to them
{\em and} (ii) they do not have any connection to a better
performing group. Therefore, $\Delta N_{\{D,y\}_{\kappa
},\{D',j\}_{\kappa }}$ in Eq.(\ref{eq4}) is given by
\begin{equation}
\Delta N_{\{D,y\}_{\kappa },\{D',j\}_{\kappa }}=N_{\{D,y\}_{\kappa
}}(1-q^{N_{\{D',j\}_{\kappa }}})q^{A_{\{D',j\}_{\kappa }}}
\label{eq7}
\end{equation}
and these agents will have the same winning probability
$S_{\{D',j\}_{\kappa}}$ as those in group $\{D',j\}_{\kappa }$ for
$t \in \{t^{\mu}_{odd}\}$. Hence the mean success rate
$\widetilde{S}_{\{D,y\}_{\kappa }}$ for agents labelled by
$\{D,y\}_{\kappa}$ for $t \in \{t^{\mu}_{odd}\}$ is given by
\begin{eqnarray}
\widetilde{S}_{\{D,y\}_{\kappa}} = \frac{1}{N_{\{D,y\}_{\kappa }}}[
\overline{N}_{\{D,y\}_{\kappa }}S_{\{D,y\}_{\kappa}} + \ \ \ \
\ \ \ \ \ \nonumber \\
\ \ \ \ \ \ \ \ \ \ \ \ \
 \sum_{D',j}\Delta N_{\{D,y\}_{\kappa },\{D',j\}_{\kappa
}}S_{\{D',j\}_{\kappa }} ]\ . \label{eq8}
\end{eqnarray}
For general $p$, Eq.(\ref{eq1}) is modified to
\begin{equation}
w^{(odd)}_{D,\kappa }=\frac{1}{N_{D}}\sum_{y=0}^{\kappa
}N_{\{D,y\}_{\kappa }}\widetilde{S}_{\{D,y\}_{\kappa }}\ .
\label{eq9}
\end{equation}
Equation (\ref{eq2}) can hence be used to evaluate the mean
success rate of agents for a given $D$, while Eq.(\ref{eq3}) gives
$\langle w \rangle$ as a function of the connectivity $p$.
Equations (\ref{eq2}) and (\ref{eq3}) coupled with
Eqs.(\ref{eq4})-(\ref{eq9}) are our main formal results.

Figure 1 shows $\langle w \rangle$ as a function of $p$. The
theory can also be used to evaluate the fluctuation $\sigma$ (see
inset). The theoretical results are in excellent agreement with
the numerical simulations. Our theory is further validated in
Fig.~2, where we compare theoretical and numerical results for the
success rates $w_{D}$ for each species-type $D$. Each $D$ species
has a distinct $p$ dependence, showing why a peak appears in the
model of Anghel {\em et al.}. For small connectivity $p$, $D=1$
and $D=2$ agents can benefit by hooking up to the better
performing $D=0$ agents. However as $p$ increases, these agents
may hook to agents belonging to groups with momentarily better
predictor performance. These links hurt the agent's success rate
since momentarily better strategies are bound to lose in
subsequent turns. Hence the success rates of $D=1$ and $D=2$
agents will increase at small $p$ and decrease at higher $p$,
while that for $D=0$ agents decreases monotonically with $p$.

Finally, having understood the underlying physics, we can propose
a performance-enhancing modification to Anghel et al.'s model.
Instead of following the best-performing agent, suppose an agent
follows the {\em second-best} performing agent among his
neighbors. Figure 3 shows that $\langle w \rangle$ is
substantially larger over a wide range of $p$. In addition, the
value of $p$ at the peak corresponds to a much larger number of
network links. Interestingly, there is recent empirical evidence
to suggest that such `second-best' rules do indeed make humans
happier on average in everyday life \cite{SciAm}.

PMH thanks the Research Grants Council of the Hong Kong SAR
Government (grant CUHK4241/01P).

%\newpage


\begin{thebibliography}{99}
\bibitem{marketimpact} For non-connected ($p=0$) agents,
$w^{(even)}$ is found by treating each agent as independent random walkers.  For
connected ($p \neq 0$) agents, the system consists of clusters of agents and
$w^{(even)}_{\kappa}$ is found by treating the clusters as independent random
walkers with different step (cluster) sizes.  The results in Figs.~1 and 2 include
this self-interaction (`market-impact') effect.
\end{thebibliography}
\end{document}